\documentclass[twocolumn,showpacs,preprintnumbers,amsmath,amssymb,
floatfix,nofootinbib]{revtex4}
\usepackage{amsmath,amsfonts,bbm,euscript,hyperref}
\usepackage{graphicx}% Include figure files
\usepackage{dcolumn}% Align table columns on decimal point
\usepackage{bm}% bold math
\usepackage{epsfig}
\epsfclipon

\newcommand{\nco}{\newcommand}
\nco{\beq}{\begin{equation}} \nco{\eeq}{\end{equation}}
\nco{\beqa}{\begin{eqnarray}} \nco{\eeqa}{\end{eqnarray}}
\def\be{\begin{equation}}
\def\ee{\end{equation}}    
\def\baray{\begin{eqnarray}}
\def\earay{\end{eqnarray}}
\def\hf{\frac{1}{2}}
\nco{\lra}{\leftrightarrow}

\nco{\sss}{\scriptscriptstyle} \nco{\dphi}{\varphi}
\nco{\lsim}{\mbox{\raisebox{-.6ex}{~$\stackrel{<}{\sim}$~}}}
\nco{\gsim}{\mbox{\raisebox{-.6ex}{~$\stackrel{>}{\sim}$~}}}

\def\IK{\relax{\rm I\kern-.20em K}}
\def\IM{\relax{\rm I\kern-.20em M}}
\def\KKLMMT{{\IK L\IM T}}
\def\pct{\%}

\begin{document}

%\preprint{McGill 04-xxx}

\title{Multibrane Inflation and Dynamical Flattening of the
Inflaton Potential}

\author{James M.\ Cline, Horace Stoica}

\affiliation{%
\centerline{Physics Department, McGill University,
3600 University Street, Montr\'eal, Qu\'ebec, Canada H3A 2T8}
e-mail: jcline@physics.mcgill.ca, stoica@physics.mcgill.ca }

\date{August, 2005}

\begin{abstract} We investigate the problem of fine tuning of the
potential in the \KKLMMT\ warped flux compactification scenario for
brane-antibrane inflation in Type IIB string theory.  We argue for
the importance of an additional parameter $\psi_0$ (approximated as
zero by \KKLMMT), namely the position of the antibrane, relative to
the equilibrium position of the brane in the absence of the
antibrane.  We show that for a range of values of a particular
combination of the K\"ahler modulus, warp factor, and $\psi_0$,
the inflaton potential can be sufficiently flat.  We point out a novel
mechanism for dynamically achieving flatness within this part of
parameter space: the presence of multiple mobile branes can lead to
a potential which initially has a metastable local minimum, but
gradually becomes flat as some of the branes tunnel out.  Eventually
the local minimum disappears and the remaining branes slowly roll
together, with assisted inflation further enhancing
the effective flatness of the potential.  With the addition of
K\"ahler and superpotential corrections, this mechanism can completely
remove the fine tuning problem of brane inflation, within large
regions of parameter space.  The model can be
falsified if future cosmic microwave background observations confirm
the hint of a large running spectral index.
\end{abstract}

\maketitle

\section{Introduction}  

The past few years have witnessed significant progress in the
building of realistic inflation models within string theory
\cite{strinf}.  The most popular approach has used the 
interaction potential between branes and antibranes as the source of
energy driving inflation; hence the interbrane separation plays the
role of the inflaton. Early attempts suffered from the assumption
that moduli  could be fixed by fiat, without specifying the
stabilization mechanism. However, explicit examples have shown that
properly fixing the moduli can interfere with the flatness which is
required for the inflaton potential \cite{KKLMMT} (\KKLMMT).  Flux
compactifications in type IIB string theory have provided a robust
way of stabilizing the dilaton and complex structure moduli of the
Calabi-Yau manifold \cite{GKP},  with the added benefit of providing
warped throats \cite{KS} which can be useful for tuning the scale of
inflation or of the standard model.  Nonperturbative effects have
been plausibly invoked to stabilize the remaining K\"ahler moduli 
\cite{KKLT,dA}. 

This last step, unfortunately, generically spoils whatever flatness
of the inflaton potential which was accomplished by warping,  and
thus requires additional fine tuning of parameters.  One can
introduce extra parameters associated with corrections to the
superpotential, which must be tuned to a part in  100 in order to
obtain the minimum amount of 60 e-foldings of inflation
\cite{KKLMMT}. Alternatively, it is possible to tune parameters
already present within the model but at the level of a part in 1000
\cite{BCSQ}.

The fine-tuning problem of the \KKLMMT\ model has been considered  by
several authors \cite{fine-tuning,CQS}. Shift symmetry of the Lagrangian has been proposed
as a possible solution \cite{shift-symmetry}.  It has also been pointed out that if Lyman
$\alpha$ constraints are viewed with skepticism, the degree of
tuning  required is considerably relaxed \cite{HH}.  Of course, this
observation also applies to other models of inflation.

In this paper we reexamine the fine-tuning issue, starting within the
context of the original \KKLMMT\ potential, without introducing
additional corrections. A closer consideration of the potential
between the brane and antibrane reveals the importance of a parameter
(which we call $\psi_0$) which was set to a certain value in \KKLMMT.
It is the distance which the brane will travel before annihilating
with the antibrane in the warped throat, assuming that the brane
starts somewhere closer to the the interior of the Calabi-Yau  at an
initial position $\psi\cong 0$.  The other important parameters
governing the shape of the inflaton potential are the warp factor of
the throat, which we denote as $\epsilon^{1/4}$, and the K\"ahler
modulus at its stabilized value, $\sigma$.  The warp factor is
tunable through the ratio of NS-NS and RR fluxes $K,M$:
$\epsilon^{1/4} = e^{-2\pi K/ 3 g_s M}$ \cite{GKP,KS}.   The VEV of
the K\"ahler modulus $\sigma$ is determined by  the nonperturbative 
superpotential $W = Ae^{-a\sigma}$ invoked by ref.\ \cite{KKLT},
where $a = 2\pi/N$ in the case of gluino condensation in an $SU(N)$
gauge theory sector which can naturally arise in these
compactifications.

If the parameter $\psi_0$ is set to zero as in \KKLMMT, then indeed
it is necessary to introduce additional sources of dependence on the
inflaton into the potential, in order to cancel the unwanted
contribution to the inflaton mass coming from the K\"ahler modulus.
However, in this work we point out that for a range of values  $0.259
< b < 0.286$ of the combination $b^2 \equiv 2\sigma
\epsilon\psi_0^{-6}$, it is possible to   achieve 60 or more
e-foldings of inflation without any additional corrections to the
potential.  Fortunately this restriction on $b$ applies only to the
simplest \KKLMMT\ model.  We will show that the restriction
disappears when one considers more general examples in which
corrections to the K\"ahler and superpotentials are taken into
account.

\begin{figure}[htbp]
\bigskip \centerline{\epsfxsize=0.4\textwidth\epsfbox{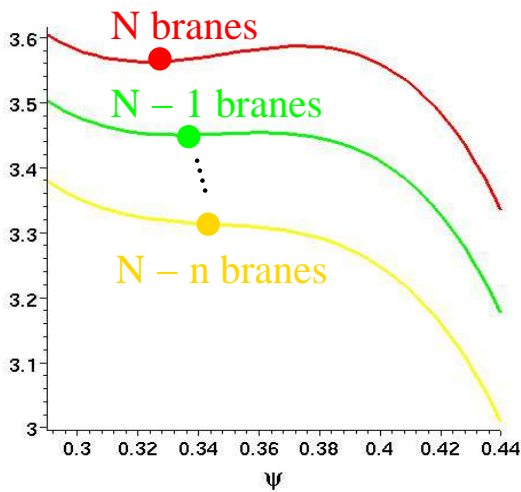}}
%\begin{verse}
%\vskip-0.25cm
\caption{Sequence of inflaton potentials due to successive
tunneling of branes toward the right.}
\label{fig1}
\end{figure}

Another crucial parameter which we consider is the number of
mobile mobiles branes, $N$,  generalizing the simplest case of only
a  single brane playing the role of the inflaton.  We refer to this
possibility as multibrane inflation. Within the aforementioned range
of $b$ values,  we observe a very interesting phenomenon, whereby
starting with a sufficiently large number $N$ dynamically leads to a
sufficiently flat inflaton potential.  At large $N$, the branes start
out being confined in a metastable minimum of the potential.  But as
successive branes tunnel out of this minimum, it becomes shallower,
until finally all the remaining branes can roll together into the
throat. This is illustrated in figure \ref{fig1}.
The potential is often very
flat when the branes start rolling, leading to a long period of
inflation.  This is a novel, dynamical mechanism for flattening the
inflaton potential, which appears to be quite specific to
brane-antibrane inflation.

%\bigskip

\section{When are Multiple Branes Allowed?}
Before looking at the dynamics of inflation with stacks of
branes and antibranes, we consider under what conditions multiple
branes are compatible with the stabilization of the extra dimensions,
{\it i.e.,} the K\"ahler modulus, $\sigma$, in supergravity language.
For each extra brane we have to introduce a corresponding
antibrane such that the tadpole cancellation condition is
statisfied. 
Adding antibranes lifts the local minimum of the potential  by an
additional amount,  given by the sum of their tensions, reduced by
the warp factor at their location in the bottom of the
Klebanov-Strassler throat.  
There is
a critical  number of antibranes that can be added such that the
potential still has a  local minimum along the $\sigma$ direction and
therefore the K\"ahler modulus  remains stable. The stronger the
warping, the larger the number of antibranes  that can be added. In
our analysis we will add antibranes such that the K\"ahler modulus
remains stabilized and heavy.  Thus it does not act like 
a second inflaton field, since it
changes very little  during the slow-rolling of the brane-antibrane
separation $\psi$.

\begin{figure}[htbp]
\includegraphics[width=0.45\textwidth,angle=0]{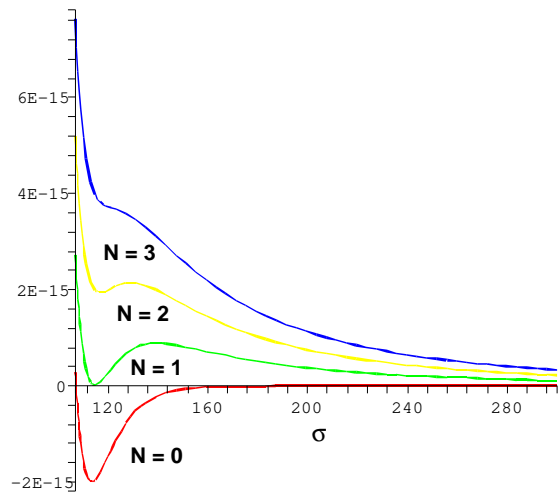}
\caption{Potential energy of the K\"ahler modulus in the 
the original KKLT scenario for $N=0,1,2$ and $3$ antibranes,
showing there can be at most 2
  antibranes without destabilizing the K\"ahler modulus.
\label{fig1n} }
\end{figure}

Not all models are suitable for constructing such a multi-brane
setup. We will give here two examples, one which works and one
which does not.  
The original KKLT \cite{KKLT} setup does not accommodate a large number of mobile
branes;    just one additional antibrane
is sufficient to lift the vacuum from anti-deSitter (AdS) to 
Minkowski (or deSitter, dS). The
barrier that prevents tha K\"ahler modulus from rolling has a height
roughly equal to the depth of the AdS minimum. Addind a second brane
lifts the potential by an amount that is roughly equal to the
height of the barrier, which destroys the local minimum.
In figure \ref{fig1n}  we take the parameters of the 
potential and the 
antibrane tension as given by \cite{KKLT}.  One sees that
adding a second brane results in a very shallow minimum and three branes
destabilizes the K\"ahler modulus. 
The potential of the K\"ahler modulus is given by 
\be
V = {aAe^{-a\sigma}\over {2\sigma^2}} \left( {1\over 3}\sigma
aA e^{-a\sigma} +  W_0 +  A e^{-a\sigma}\right)+{ND\over \sigma^3}
\ee
with  $W_0 =- 10^{-4}$, $A=1$, $a =0.1$ $D=3 \times 10^{-9}$. 
\vskip-0.25cm

\begin{figure}[htbp]
\includegraphics[width=0.45\textwidth,angle=0]{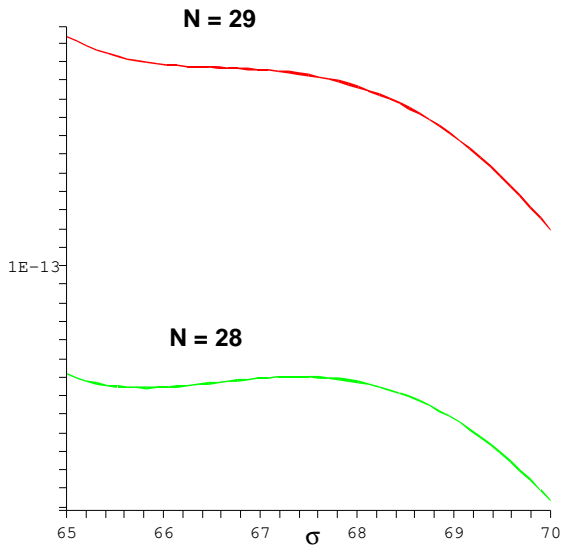}
\caption{Using
  a racetrack superpotential and a throat with stronger warping one
  can add a large number (28 in this example) of antibranes 
while keeping the K\"ahler modulus
  stable.\label{fig2n}}
\end{figure}

The situation is dramatically improved if one uses a racetrack
superpotential as in ref.\ \cite{KL}. The minimum is already
Minkowski and adding antibranes in a highly warped throat
lifts the vacuum to dS without destabilizing the K\"ahler modulus.
The potential is:
\baray
&&V= {e^{-2(a+b)\sigma}\over 6\sigma^2}(b B e^{a\sigma}+ a A
e^{b\sigma})
\\ 
&\times& \left
[Be^{a\sigma}(3+b\sigma)+e^{b\sigma}(A(3+a\sigma)+3e^{a\sigma}W_0)\right]
+\frac{ND}{\sigma^3}\nonumber
\earay
where we choose the same parameters as in \cite{KL},
$A=1,\ B=-1.03,\ a=2\pi/100,\ b=2\pi/99,\ W_0= -2\times 10^{-4}$
and $D = 10^{-9}$. With these values the critical number of branes
turns out to be $N = 28$, illustrated in fig.\ \ref{fig2n}

In the remaining analysis, we will not be concerned with the
particular model of K\"ahler modulus stabilization, beyond the
assumption that it admits the required number of additional branes.
It will also be found that a large number is not required, so the
racetrack model mentioned here is sufficient to establish the
existence of a viable model.

\section{Inflaton Action}

Our starting point for the dynamics of inflation 
is the Lagrangian for the mobile brane position
$\psi$ in the \KKLMMT\ framework.
To simplify the analysis, we will assume that all moduli (notably the
K\"ahler modulus $\sigma$) are heavy except for the inflaton, and that the
4D effective cosmological constant vanishes at the end of inflation.
The Lagrangian can be parametrized as
\beq
\label{Lag}
{\cal L} =  {6\sigma\over r^2}\dot\psi^2 - 
{T\epsilon\over r^2}\left(1 +
	{\epsilon\over(\psi-\psi_0)^4}\right)^{-1}
\eeq
where $r = (2\sigma- \psi^2)$ \cite{dWG}, 
$T$ is related to the tension $\tau$ of the 3-brane or antibrane
by $T = 2\tau/g_s^4$ \cite{KKLT}, $\epsilon$ is
the warp factor, $\psi$ is the 3-brane position, and
$\psi_0$ is the location of the antibrane. To keep our work
self-contained, this expression is
derived in Appendix A, in just the same way as in \KKLMMT\, except
for the omission of two approximations:  (1) we do not assume that
$\psi \gg \psi_0$, and (2) we do not Taylor-expand the term
$(1 +	{\epsilon/(\psi-\psi_0)^4})^{-1}$.

Let us first comment on point (1) above.
In \KKLMMT\ and certain other investigations of brane-antibrane
inflation, it was assumed that inflation takes place sufficiently far
from the antibrane to justify the approximation $|\psi - \psi_0|
\cong \psi_0$.  On the other hand, the $\psi_0$ parameter was kept
in  \cite{BCSQ} and \cite{IT}.  Physically, it may be  
important to allow for nonzero $\psi_0$ because of the possibility
that the $\psi$-dependence coming from the $(2\sigma- \psi^2)^2$
factor in the potential will tend to stabilize the brane at some
point other than the bottom of the warped throat, in the absence of
the brane-antibrane interaction.  An example where this is
particularly clear is that of two symmetrically placed warped
throats, each with its own antibrane \cite{IT}.  In that case the
potential (translated to our notation) was taken to be
\beq
V = {T\epsilon\over r^2}\left(2 - 
	{\epsilon\over(\psi-\psi_0)^4  }
- 	{\epsilon\over(\psi+\psi_0)^4}
\right)
\eeq
From symmetry considerations it is obvious in this model that there
exists a metastable or unstable equilibrium point exactly midway
between the two throats, at $\psi=0$, which cannot coincide with the
bottom of either throat.  If we simply remove one of the throats, the
result corresponds to our starting point (\ref{Lag}).  

The presence or absence of the $\psi_0$ parameter explains
differences in the amount of required fine-tuning of the potential
found in different investigations.   To elucidate this, recall that
the $\psi$-dependence in the volume modulus $r=(2\sigma- \psi^2)$ is
the origin of the $\eta$ problem pointed out in \KKLMMT---the mass
of the inflaton is of order $\sqrt{V}\sim H$ in Planck units, ruining the
slow roll condition.  It was noticed by \cite{BCSQ} that it is
possible to tune parameters of a model with a potential similar to
that in 
(\ref{Lag}) so that the slow-roll conditions are satisfied, even
without adding any extra $\psi$-dependent corrections to the
superpotential, by varying only the parameters which already
appear.   However, the degree of  tuning needed was more severe than
in \KKLMMT\, where corrections to the superpotential were invoked.  One
way of understanding this is to realize that if $\psi_0=0$ and only
corrections of order $\psi^2$ are added, then the potential
automatically has $V'(0)=0$, so that one of  the slow roll parameters
$V'/V$ is small without any tuning.  Then only the $\eta$ parameter
($V''/V$) needs to be tuned.  In our potential however, both 
slow-roll parameters must be tuned, for generic parameters in the
potential.  In \cite{BCSQ} it was found that
60 e-foldings of inflation could be obtained by tuning the warp
factor to one part in 2000, whereas \KKLMMT\ only needed one part in
100 tuning.  Although this sounds contrary to the spirit of the
present paper, it was due to the fact that no systematic study of the
effect of varying $\psi_0$ was done in \cite{BCSQ}; instead only one
fixed value was considered.

We note that the behavior of the potential as $\psi\to \psi_0$ is
classically correct, since by derivation it is of the form
$1/h(\psi)$ where $h(\psi)$ satisfies $\nabla^2 h =
C\delta(\psi-\psi_0)$ in the 6D Calabi-Yau manifold.  Although the
full brane position must be specified by 3 complex $\psi$
coordinates, we are keeping the angular ones fixed and following
only the radial coordinate for  the brane-antibrane separation.

We also comment on our choice of avoiding the second
approximation (point (2) above), by not Taylor expanding $(1 +
{\epsilon/(\psi-\psi_0)^4})^{-1}$.  We choose not to do so
because there is no advantage in making this expansion---it
only leads to an unphysical divergence in the potential energy as
the brane-antibrane distance shrinks.  The unexpanded form has good
behavior as $\psi\to\psi_0$, including the vanishing of the potential
at the minimum.  

\subsection{Multibrane Lagrangian}
We will be interested below in the generalization to more than
one brane rolling.  The Lagrangian for $N$ branes and $N$ antibranes
is
\beq
\label{LagM}
{\cal L} =  {6\sigma\over r^2}\sum_{i=1}^N\dot\psi_i^2 - 
{NT\epsilon\over r^2}\left(1 +
	\sum_{i=1}^N{\epsilon\over(\psi_i-\psi_0)^4}\right)^{-1}
\eeq
where now the volume modulus is 
\beq
	r = 2\sigma- \sum_{i=1}^N\psi_i^2
\eeq
Later we will specialize to the situation where all the branes
are coincident, $\psi_i = \psi$.  Then the potential takes the form
$N\epsilon Tr^{-2}(1 + N\epsilon(\psi-\psi_0)^{-4})^{-1}$.

\section{Conditions for Inflation}

The optimal potential for achieving slow roll inflation is one where
both the first and second derivatives vanish simultaneously at some
critical point $\psi_c$.  Although this requires fine tuning, we will
discuss a mechanism where the tuning can automatically occur through
a dynamical process.  To characterize the tuning,
we introduce new parameters
\beqa
	a &=& {N\epsilon\over\psi_0^4},
	\quad b^2 = {2\sigma\epsilon\over\psi_0^6}, \quad
	x = {\psi\over \psi_0}
\eeqa
such that the potential is proportional to
\beq
\label{hatV}
	\hat V(x) = {1\over \left({b^2/ a} - x^2\right)^{2}}
		\,\left(1 + {a\over (x-1)^{4}}\right)^{-1} 
\eeq
The conditions for the potential to have a flat point, as defined
above, are
%\beq
%\label{crit}
%	x( (1-x)^5 + a ) = b^2,\quad a = 5(1-x)^4 - 6 (1-x)^5
%\eeq

\beq
\label{crit}
a = (6x-1)(x-1)^4, \quad 5x^2(x-1)^4 = b^2
\eeq

These equations can be solved analytically, as the solution of a
cubic equation, but the expression is not very enlightening.  Instead
we give an approximate description, by first
considering $b$ as a free parameter, and determining 
as a function of $b$ 
the critical values $a_c$ and $x_c$ which satisfy (\ref{crit}).
For consistency, we should also require that $b^2 \ge a$; otherwise the
potential blows up at $x = b/\sqrt{a} < 1$, before the brane reaches
the antibrane.  The range of values where $a_c > 0$ (obviously also 
required for a
physically sensible solution) and $a_c < b^2$ turns out to be rather
narrow, 
\beq
\label{range}
0.259 < b < 0.286,
\eeq
and in this region $a_c$ is to a good approximation a
linear function of $b$:
\beq
\label{ac}
	a_c \cong 3b - 0.77
\eeq
One can also approximate the critical value $x_c$ by a quadratic
form, $x_c \cong -0.146 + 1.21 b + 6(b-0.259)(b-0.286)$.

We will show later that the narrowness of the interval (\ref{range})
is an accident, in the sense that physically well-motivated
perturbations to the potential can remove this restriction. 
Therefore we will ultimately not regard it as a fine tuning.  However
the special value $a_c$ for a given value of $b$ clearly does
represent a fine tuning of the parameters. It is this condition which
will be addressed by a dynamical mechanism.  Namely, by starting with
a large enough number $N$ of branes, one can initially have $a > a_c$
(but still $a < b^2$).  The potential always has a metastable minimum
from which branes can tunnel in this case. After each tunneling
event, the value of $a$ diminishes by a discrete amount. At some
critical number of branes, $N_c$, the curvature changes so that the
branes roll instead of being trapped.  It becomes a quantitative
question, depending on the values of $\sigma$ and $\epsilon$, whether
the potential is sufficiently flat to give enough inflation.  We will
show that in fact it is almost always flat enough to give at least
60 e-foldings of inflation.

To find out how likely it is that adequate inflation will occur,
we have explored the parameter space $\epsilon,\ \sigma,\ b$, subject
to the aforementioned restrictions and the assumption that $N$ starts
out being larger than the critical value, where all the branes start
to roll.   For each value of $\{\epsilon,\ \sigma,\ b\}$, the
quantities $\psi_0$ and
$N_c$ are determined.  The fields start from rest at the value
where the potential has a local minimum with $N=N_c+1$
branes, on the assumption that the nontunneling branes do not move
during the tunneling event.  The equations of motion are
\beq
	{d\pi_i\over dt} = -3H\pi_i - 
	{\partial V\over \partial \psi_i},\quad {d N_e\over dt} = H
\eeq
for the canonical momentum 
\beq
	\pi_i = {12\sigma\over r^2}\dot\psi_i
\eeq
and number of e-foldings $N_e\equiv \ln a(t)$, where the Hubble parameter is given by
$H^2 = (6\sigma N\dot\psi^2/r^2 + V)/(3M_p^2)$.
Since all the fields move together, we need consider only one 
equation for the inflatons, which we do numerically.

\begin{figure}
\centerline{\epsfxsize=0.5\textwidth\epsfbox{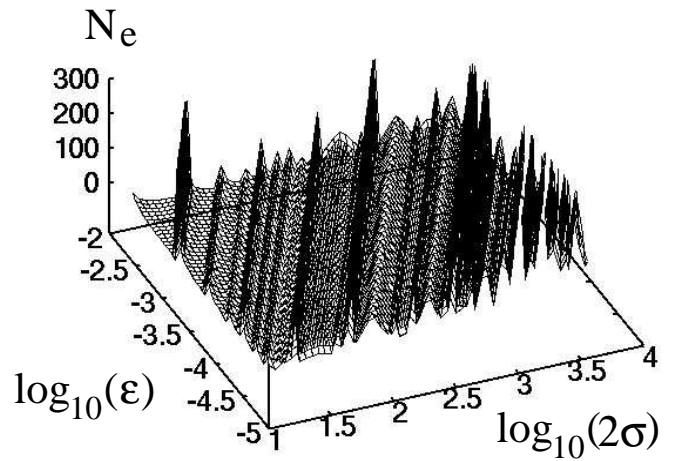}}
%\begin{verse}
%\vskip-0.25cm
\caption{%\small\noindent Figure 2. 
$N_e$ versus $\log_{10}\epsilon$ and $\log_{10}2\sigma$ for $b=0.27$,
in the range $N_e < 300$.}
\label{fig2}
\end{figure}

The result is that, for $b$ satisfying (\ref{range}), there is a high
probability of getting enough inflation, for random values of
$\epsilon, \sigma$ selected from within a reasonable range, 
$-5 \le \log_{10}\epsilon \le -2.4$, $1 \le \log_{10}2\sigma \le
3.6$.  Scanning uniformly over this region, we find that 
$85\pct$ of tries result in more than 55 e-foldings of
inflation when $b=0.27$, and an even greater percentage occurs
for other values of $b$. The results are
illustrated for $b=0.27$ in Figure \ref{fig2}, showing the number of
e-foldings, $N_e$ (omitting values where $N_e > 300$ for clarity) 
 versus $\log_{10}\epsilon$ and $\log_{10}2\sigma$.

In figure \ref{fig2} it can be seen that $N_e$ is constant along lines of constant
$A \equiv (2\sigma)^{2/3}\epsilon^{-1/3}$.  This can be understood by rewriting the
potential in terms of the canonically normalized field $\phi =
\psi_i\sqrt{3N/\sigma}$:
\beq
	V = {T\,N\over A^3}
	 \left( 1 - \frac 16\phi^2\right)^{-2}
 	\left(1 - {36(N/A)^3\over(\phi-\phi_0)^{4}}\right)^{-1} 
\eeq
where $\phi_0 = b^{-1/3}\sqrt{3N/A}$.  Thus for fixed $b$, 
the shape of the potential, power spectrum, and duration of inflation depend only upon 
combination of parameters in $A$, and not on the orthogonal
combination $B = (2\sigma)^{1/3}\epsilon^{2/3}$. 
We can therefore restrict our exploration of parameter space to a one-dimensional
scan along the direction of $A$.  Moreover, we can show that the
number of branes $N$ depends in a very simple way on the
value of $A$: $N = a b^{-4/3} A$.
This follows from eliminating $\psi_0$ from the definitions of $a$
and $b$.  Therefore the critical number of branes is given by
\beq
\label{Nc}
	N_c = a_c\, b^{-4/3}\, {(2\sigma)^{2/3}\over \epsilon^{1/3}}
\eeq
and for a fixed value of $b$, we can treat the parameters $A$ and
$N_c$ interchangeably.
Of course (\ref{Nc}) can't be satisfied exactly, because $N_c$
has to be an integer.  The accidental discrepancies between the
optimal value of $N_c$ and the closest integer below this value 
will affect how long inflation actually lasts.

\begin{figure}
\centerline{\epsfxsize=0.5\textwidth\epsfbox{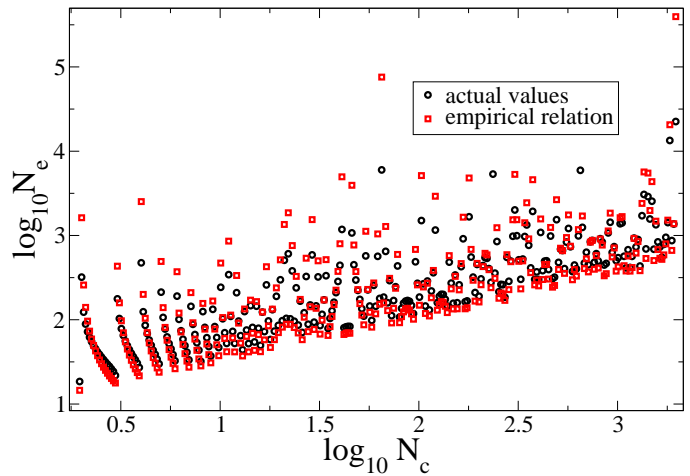}}
%\begin{verse}
%\vskip-0.25cm
\caption{%\small\noindent Figure 3. 
$\log N_e$ versus the critical number of branes $\log N_c$, $b=0.27$.
The empirical relationship (\ref{emp}) is also plotted.}
\label{fig3}
\end{figure}

In figure \ref{fig3} we show the correlation between number of e-foldings
and the critical number of branes, scanning uniformly in the
interval $1 < \log_{10}A < 4$ of parameter space.  Although the two quantities
are generally correlated, there are order-of-magnitude variations
in $N_e$ over narrow ranges of $N_c$.  Figure \ref{fig3} demonstrates that
both features can be
matched fairly well by an empirical relationship.  
Let $\Delta N_c$ be the
discrepancy between the ideal value of $N_c$ and the closest integer
below this value.  Then we find that
\beq
\label{emp}
	N_e \cong 10 {\sqrt{N_c}\over \Delta N_c}
\eeq
The factor $1/\Delta N_c$, is understandable
since if $\Delta N_c=0$, the potential is
tuned to be perfectly flat at some position $\psi=\psi_c$.

 The factor 
$\sqrt{N_c}$ in (\ref{emp}) can be partly understood in terms
 of assisted inflation \cite{assisted}.  With $N$ fields rolling
simultaneously, the kinetic term is multiplied by $N$ relative to 
one field.  We should rescale the fields by $\psi \to\psi/\sqrt{N}$,
and the slow-roll parameters by $1/N$.  One therefore expects an
enhancement in the length of inflation by some power of
$N_c$. It would of course be satisfying to understand why
that power is $1/2$ in the present model.  We have not yet succeeded
in analytically deriving this result.  However it is not
surprising that $N_e$ fails to be enhanced by a full factor
of $N_c$ as would be the case for ordinary assisted inflation, because
the potential contains $N\psi^2$ in the K\"ahler modulus.  Only if
the potential was independent of $N$ would we be sure that
 $N_e \sim N_c$.  In that case, transforming to the canonically
normalized field would cause $V(\psi) \to V(\phi/\sqrt{N})$ and
the slow roll parameters would go like
$\eta \sim V''/V \sim 1/N$.  The surprise is that there is any assisted 
inflation effect at all.  The reason for it is that at the critical
value of $N$, the potential is flatter than it would generically be
for an arbitrary function of $N\psi^2$.

\begin{figure}
\centerline{\epsfxsize=0.5\textwidth\epsfbox{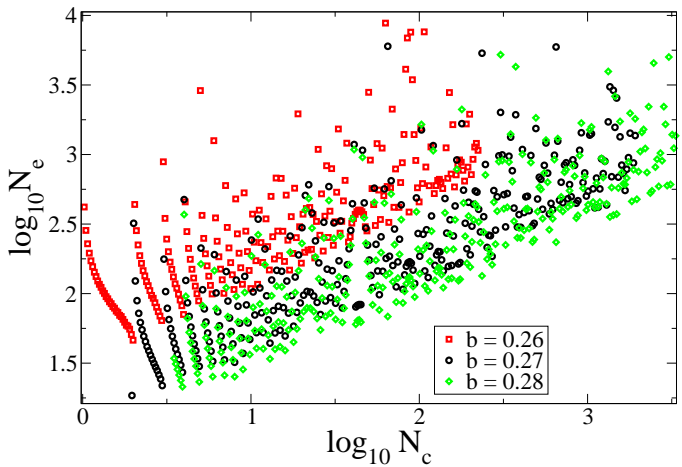}}
%\begin{verse}
%\vskip-0.25cm
\caption{%\small\noindent Figure 4. 
$\log N_e$ versus $\log N_c$ for $b=0.26,\ 0.27,$ and $0.28$.}
\label{fig4}
\end{figure}

The above results were for $b=0.27$.  Allowing $b$ to vary, ones
finds that smaller $b$ tends to give more inflation with a smaller
number of branes.  This dependence is shown in figure \ref{fig4}.  The
main observation, though, is that the great majority of all cases
yield more than $60$ e-foldings of inflation.

\section{Scalar Power Spectrum}
\begin{figure}
\centerline{\epsfxsize=0.5\textwidth\epsfbox{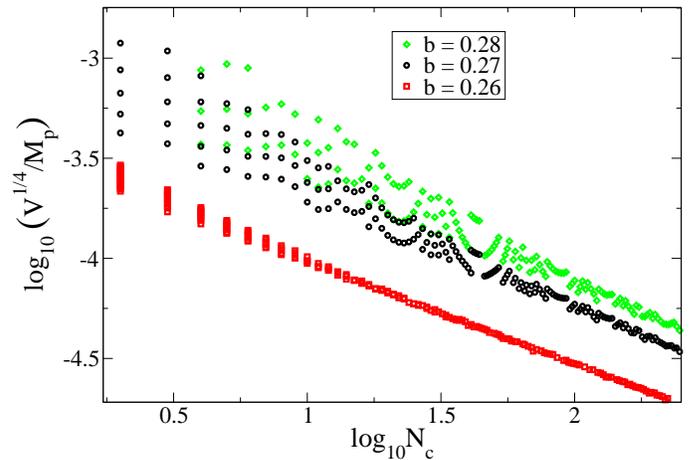}}
%\vskip-2cm
\caption{%\small\noindent Figure 5. 
Log of the inflationary scale as a function of the number of branes.}
\label{fig5}
\end{figure}

In addition to having a long enough period of inflation, we must
obtain a spectrum of density perturbations in agreement with
observational constraints from the Cosmic Microwave Background.
To compute the power spectrum, we use the following expression 
\cite{SS} which
applies to a set of noncanonically normalized fields with Lagrangian
${\cal L} = \frac12 G_{ij}\dot\psi^i\dot\psi^j - V$:
\beq %\delta_{\sss H}^2
	P(k) = {V\over 75\pi^2 M_p^2} 
	{\partial N_{e}\over \partial\psi^i} 
	{\partial N_{e}\over \partial\psi^j} G^{ij} = 
	{N_c V H^2\over 75\pi^2 M_p^2}\, {r\over 6\dot\psi^2}
\eeq
These quantities are evaluated at horizon crossing of the relevant
wave number, when $k=a H$.  To normalize the spectrum, we evaluate
$P(k)$ for modes which crossed the horizon 55 e-foldings before
inflation ended, corresponding to a scale of inflation around 
$V^{1/4} \sim 4\times 10^{14}$ GeV.  The consistency of this
assumption will be verified. The COBE normalization implies that
$P\cong 4\times 10^{-10}$ for these modes. This allows us to normalize the
scale of the potential by fixing the value of the parameter $T$
which is proportional to the 
unwarped brane tension.  
The results are expressed in terms of the scale of inflation
relative to the Planck scale.  In figure \ref{fig5} we plot $\log(V^{1/4}/M_p)$
as a function of the number of branes.  (In the previous section it
was explained that, just like
$N_e$, spectral properties depend only on the same combination of
$\sigma$ and $\epsilon$ which determine $N_c$.)  From the figure one
sees that the inflationary scale ranges between $10^{-3}$ and
$10^{-4.5} M_p$, roughly consistent with our assumption of $\sim 55$
e-foldings of inflation.  The string scale, which determines the
brane tension, must be in about the same range.  If the number of
branes is small enough, the inflationary
scale can be sufficiently high for the tensor contribution 
to the cosmic
microwave background to be observable in future experiments.

\begin{figure}
\centerline{\epsfxsize=0.5\textwidth\epsfbox{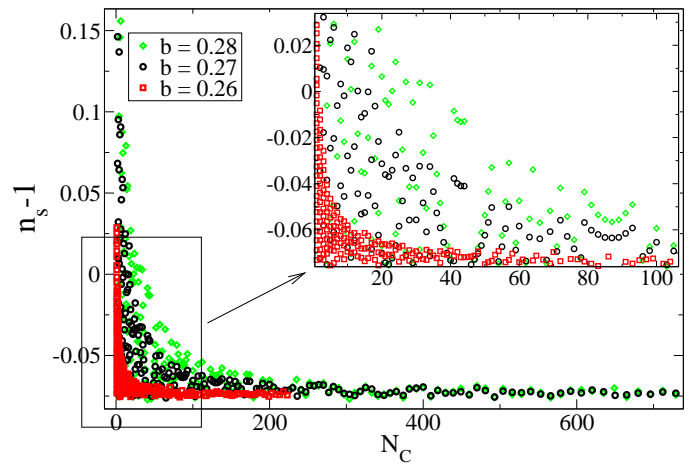}}
%\vskip-2cm
\caption{%\small\noindent Figure 6. 
Deviation of scalar spectral index from unity versus $N_c$}
\label{fig6}
\end{figure}

The scalar spectral index is given by $n_s = 1 + d\ln P/d\ln k$.  We
have numerically evaluated it over the relevant range of parameters.
Like the other quantities, it is most conveniently displayed as a
function of the number of branes.  The deviation $n_s-1$ at 55
e-foldings before the end of inflation is shown as a function of
$N_c$ in figure \ref{fig6}.  There is a wide range of possible values,
$-0.07 < n_s -1 < 0.15$.

\begin{figure}
\centerline{\epsfxsize=0.5\textwidth\epsfbox{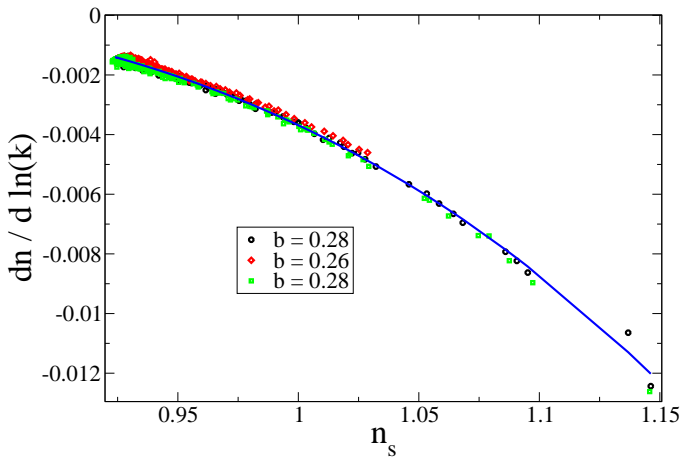}}
\nobreak
\caption{%\small\noindent Figure 7. 
$dn/d\ln k$ versus $n_s$.}
\label{fig7}
\end{figure}

It is also interesting to consider the running of the spectral index,
$dn/d\ln k = d^2\ln P/d\ln k^2$, for which WMAP has given a hint of
a larger-than-expected negative value, $dn/d\ln k  \cong -0.02$
\cite{WMAP-inf}. 
Although the error bars are large enough to be consistent with no
running, the suggestion of large running has attracted attention.
Generically, within the slow-roll formalism,  one expects 
$dn/ d\ln k \sim (n_s - 1)^2$, 
which is usually smaller than the measured central value. 
Our numerical survey shows that it is possible to have
rare instances of running as negative as $-0.01$, even though inflation
lasts long enough.  However, as shown in
figure \ref{fig7}, it is still true that  $dn/ d\ln k \sim (n_s - 1)^2$.
The rare cases of large running correspond to occasional extreme
values of the spectral index.

We have checked that the spectral index and running are in good
agreement with the predictions of slow roll inflation,
$n-1 = 2\eta-6\varepsilon$, $dn/d\ln k = -2\xi +
16\epsilon\eta-24\varepsilon^2$, 
where $\varepsilon = \frac12(V'/V)^2$,
$\eta = V''/V$ and $\xi = V'V'''/V^2$, in units where $M_p=1$.
The derivatives must be taken with respect to the canonically
normalized fields, $\phi_i = (r/\sqrt{12\sigma})\psi_i$.
In all cases, $\varepsilon\sim 10^{-6} - 10^{-9}$, whereas 
$\eta\sim -0.03$ and $\xi\sim 10^{-3}$.  The running is completely
dominated by $\xi$.

\begin{figure}
\centerline{\epsfxsize=0.5\textwidth\epsfbox{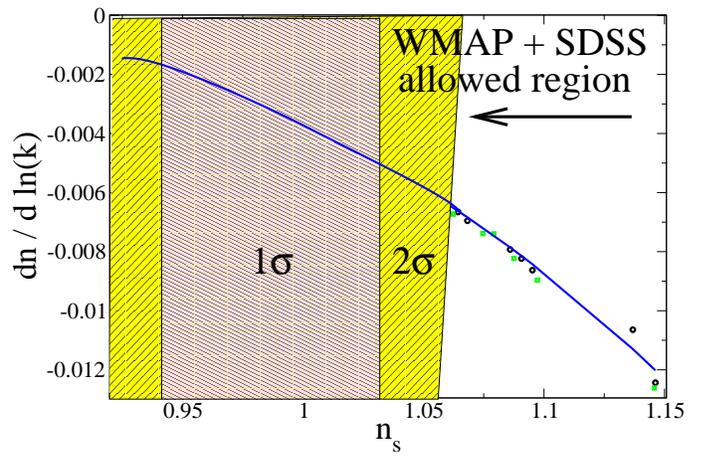}}
\nobreak
\caption{%\small\noindent Figure 8. 
$dn/d\ln k$ with $n-1$, with WMAP$+$SDSS constraints shown.}
\label{fig8}
\end{figure}

It is interesting to compare the predicted values of the running
and the spectral index with experimental constraints from combined
WMAP and Sloan Digital Sky Survey Lyman $\alpha$ data \cite{Seljak}  
The allowed region is superimposed on the predictions in figure
\ref{fig8}.  Although most of the predicted points are allowed, we note that
the few examples with large running have too large a deviation of
$n_s$ from unity.  Thus if large running should be experimentally
confirmed in the future, it will rule out the models under
consideration.

\section{Generalized Models}

The previous analysis shows that within a narrow range of the 
parameter $b =  \sqrt{2\sigma\epsilon}/\psi_0^3$, a long period of
inflation will always ensue, as long as the initial number of mobile
branes exceeded the critical number $N_c$. But since $b$ looks
fine-tuned, this is not a complete solution to the
problem of naturalness for the inflaton potential.  Could this
be an accident of the particular form of the simplest \KKLMMT\
potential?  To answer this, we explore
the effect of perturbing the model by corrections which are expected
to be present, but which we have so far ignored for the sake of
simplicity.

At least two kinds of modifications to the basic \KKLMMT\ model are
generically expected: corrections to the superpotential and to the
K\"ahler potential.  
A simple example of the former was considered in \cite{KKLMMT},
where the superpotential for the K\"ahler modulus $T = \sigma +
i\theta$ and inflaton was taken to be 
\beq
	W = W_0 +g(T)(1 + \delta\psi^2)
\eeq
leading to a contribution to the potential of the form 
\beq
\label{deltaV}
	\delta V = {\alpha + \beta\psi^2 + \gamma\psi^4
	\over (2\sigma - \psi^2)^2 }
\eeq
with $\sigma$-dependent coefficients $\alpha,\ \beta,\ \gamma$.  This
should be added to the brane-antibrane potential which we have been
using so far.  In the case of $N$ branes, we take 
$W = W_0 +g(T)(1 + \delta\sum_i\psi_i^2)$.
To simplify the analysis, we consider examples
where $\alpha,\ \beta,\ \gamma$ are chosen in such a way that the
full potential continues to have a global minimum at $\psi=\psi_0$,
where $V(\psi_0)$ is assumed to vanish:
\beq
\delta V \sim N\delta{(\psi^2 - \psi_0^2)^2\over (2\sigma - N\psi^2)^2}
\eeq
We have included the factors of $N$ appropriate to the multibrane
case, when all branes are coincident.

The second kind of correction involves the K\"ahler potential, which
for $N$ branes was taken to be 
$K = - 3\ln(T+\bar T - \sum_i|\psi_i|^2)$ \cite{dWG}.
This was always understood to be an approximate expression, good when
$\psi_i^2 \ll \sigma$, but subject to higher order corrections in
$\psi_i$,  for example $|\psi_i|^4$.  The
denominator in $V + \delta V$ would then be replaced by
\beq
	(2\sigma - N\psi^2)^2 \to (2\sigma - N(\psi^2 + d\psi^4))^2
\eeq
As shown in Appendix B, the K\"ahler corrections also give contributions to the numerator
of the F-term potential, whose Taylor series starts with a term of
order $\psi^4$.  For small $\psi$, we can absorb this leading term
into the $\gamma$ coefficient of (\ref{deltaV}).  The kinetic term
${\cal K}_{a\bar b}\dot\phi^a\dot{\bar\phi}^b$ 
also gets ${\cal O}(d\psi^2)$ corrections, which can be inferred from eq.\
(\ref{Kmetric}), but these are negligible whenever the branes are
slowly rolling.

\begin{figure}
\centerline{\epsfxsize=0.5\textwidth\epsfbox{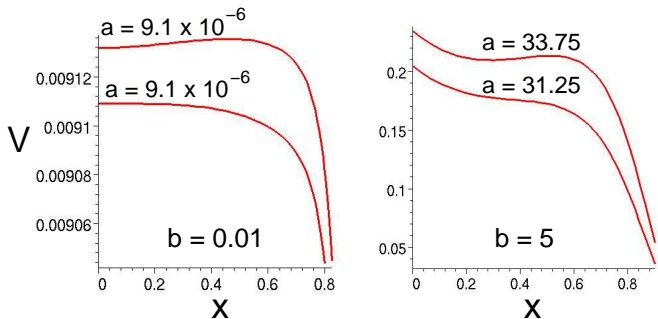}}
\nobreak
\caption{%\small\noindent Figure 9. 
Rescaled brane-antibrane potential (\ref{newhatV}) including K\"ahler and superpotential
corrections.}
\label{fig9}
\end{figure}

By the above reasoning, the two kinds of corrections enter into the
rescaled potential (\ref{hatV}) as
\beq
\label{newhatV}
	\hat V \to {1\over \left({b^2\over a} - x^2 - dx^4\right)^{2}}
		\,\left( c(x^2-1)^2 + \left(1 + {a\over (x-1)^{4}}\right)^{-1}
\right)
\eeq
By experimenting with this generalized potential, one finds that the
new coefficient $c$ allows for the removal of the lower bound on
$b$ in (\ref{range}).  Moreover, negative values of $d$ can remove
the upper bound.  It is easy to find values of $c$ and $d$ for which
the potential has the desired behavior as a function of $N$, as in
figure \ref{fig1}, regardless of the value of $b$.  For example $c=0.1$, 
$d=-0.5$ yields the potentials shown in figure \ref{fig9}, at $b=0.01$ and
$b=5$.
It is important that
the new parameters $c$ and $d$ are independent of the number of
branes.  Thus $a$ continues to be the only parameter which changes
due to a brane tunneling event.

\begin{figure}
\centerline{\epsfxsize=0.5\textwidth\epsfbox{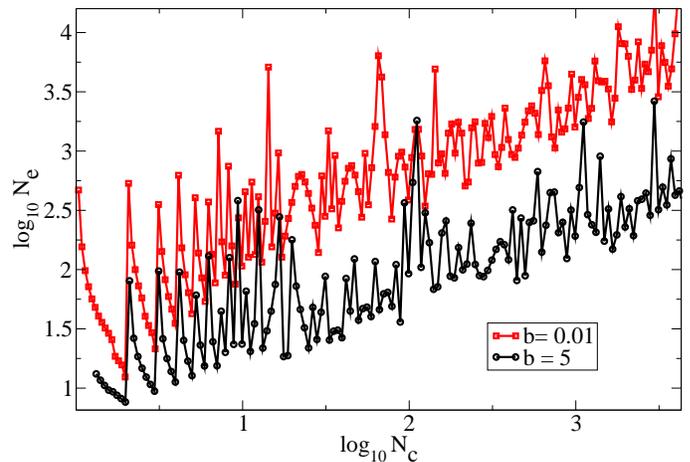}}
\nobreak
\caption{%\small\noindent Figure 10. 
Log of number of e-foldings versus log of the critical number of branes, for
the examples illustrated in figure 9.}
\label{fig10}
\end{figure}

As in the previous section, we scanned over parameters for which
$A$ varies, hence the number of branes, while keeping 
$b,\ c,\ d$ fixed, for the two examples with $b=0.01$ and $b=5$.
Figure \ref{fig10} shows the number of e-foldings versus number of branes.
The general growth of $N_e\sim\sqrt{N_c}$ is evident, as it was
for the simple model of the previous section.

\begin{figure}
\centerline{\epsfxsize=0.5\textwidth\epsfbox{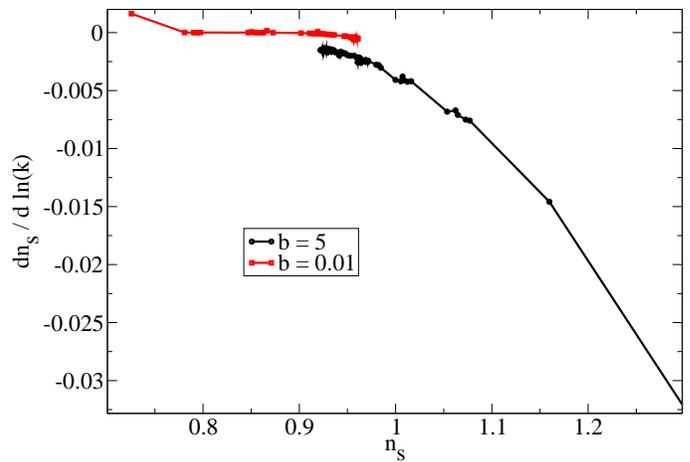}}
\nobreak
\caption{%\small\noindent Figure 11. 
$dn/d\ln k$ versus $n_s$ for the
models with K\"ahler and superpotential corrections.
}
\label{fig11}
\end{figure}

Figure \ref{fig11} shows the correlation of $dn/d\ln k$ with $n_s$.
It shows that the small-$b$ model tends to have a
red spectrum, with $n_s < 1$, and  negligible running.
The large-$b$ model also has $n_s < 1$ for most realizations, but
for rare cases with $N\lsim 20$, it is possible to get large
deviations from a flat spectrum with $n_s > 1$.  
These are accompanied by sizable running, but as in figure
\ref{fig8}, such points occur outside of the experimentally allowed region.
The overall conclusion is that the generalized models have similar
qualitative behavior to the simplest class which was studied in
sections III-IV.

\section{Caveats and Conclusions}  

In this work we have presented a very general means whereby a long
period of inflation, with a suitably flat spectrum of density
perturbations,  can be automatically obtained after tunneling of
branes, from a metastable position in a Calabi-Yau manifold, into a
warped throat containing antibranes. It depends only on the fact that
the brane-antibrane potential generically has such a metastable
minimum when the number of branes is sufficiently large, and the 
possibility of starting at a place in the string theory landscape
where the number of mobile branes exceeds the threshold for having a
minimum.  Once a certain critical number of branes $N_c$ is reached,
the minimum disappears and the branes start rolling slowly to the
true minimum.  They automatically start from the same position, which
is guaranteed to be close to the flattened region of the
potential.

In this scenario, an arbitrarily long time will be spent in  a
metastable de Sitter phase resembling old inflation, before the slow
rolling period (similarly to ref.\ \cite{new-old}).  Each tunneling event results in a small bubble of
open universe, within which another period of quasi-de Sitter
expansion will occur.  The process repeats itself until the critical
number of branes is reached.   For us, these long periods of old
inflation are irrelevant---we have focused entirely on the final
phase of conventional inflation associated with the smooth motion of
the branes until they annihilate with antibranes in the throat,
hopefully leading to the reheating of the universe  \cite{reheating}, and perhaps the
production of cosmic superstrings \cite{cosmic-string}.

We numerically scanned over large ranges of the logarithm of relevant
combinations of parameters which govern the shape of the inflaton
potential, and found that large fractions ($ \sim 10 - 90\%
$) of the trials resulted in inflation with no need for tuning,
whenever the flattening mechanism could occur.  This seems to be 
a quite promising approach for ameliorating the need for fine-tuning
within the framework of the string-theory landscape \cite{landscape}.

For convenience of numerical integration, we chose a form of the
potential which has good behavior as the branes reach the bottom of
the throat.   This form, which differs from the commonly used
Taylor-expanded form, is in fact the result of a direct perturbative
computation, and does  not require any special assumptions.  However,
the perturbative derivation does assume that
$N\epsilon/(\psi-\psi_0)^4 \ll 1$ in order to be valid.  One might
therefore wonder how sensitive our results are to the behavior of the
potential when this approximation is breaking down.  We have 
investigated what happens if the Taylor expansion  $1 -
N\epsilon/(\psi-\psi_0)^4$ is used in place of  $(1 +
N\epsilon/(\psi-\psi_0)^4)^{-1}$, and we also checked different ways
of regulating the singularity, like  $1 - N\epsilon/(N\epsilon +
(\psi-\psi_0)^4)$.  For all these alternatives, we found that
although the specific parameter values where a flat potential occurs
may be somewhat sensitive to the behavior of the potential near
$\psi_0$, the mechanism of multibrane inflation nevertheless works in
the same way as we have presented.  We therefore believe it to be a
robust result, which is qualitatively quite different from field
theory models of inflation. 

Although the dynamical flattening of the potential requires a rather
special range of parameters in the simplest \KKLMMT\ model, we showed
that this restriction can be lifted once corrections to the
superpotential and the K\"ahler potential are taken into account.  
We exhibited an example of such corrections that works, without
systematically exploring the parameter space of such corrections.
Subjectively, it seems that the parameter space (referred to as 
$\{b,c\}$ in section V) where multibrane inflation works will be
large, given that it was easy to find a successful example,  but a
more complete investigation could be made.  Moreover, such
corrections should in principle be computed from a fully string
theoretic starting point, which might restrict the allowed ranges
of $\{b,c\}$.

Another issue which deserves further study is the precise
relationship between the radial coordinate $r$ of the brane within
the warped throat, and the field $\psi$ which is used to describe the
inflaton position in the supergravity description.  The exact 
correspondence between these fields was not needed in \KKLMMT\
because the main dependence on $\psi$ was taken to be through the
K\"ahler modulus factor (and superpotential corrections), rather than
a combination of these with the brane-antibrane potential.  We have
simply identified the two fields in our analysis, but the actual
relation could be more complicated, possibly introducing additional
parameters into the Lagrangian.

\section*{Acknowledgments}
We thank C.\ Burgess and F.\ Quevedo 
for collaboration during the early stages of
this work.  
We also thank H.\ Firouzjahi, S.\ Kachru, R.\ Kallosh, L.\ Kofman, 
A.\ Linde,  I.\ Neupane,
E.\ Silverstein, L.\ Susskind,  and  H.\ Tye for helpful comments
and discussions.  JC and HS are supported by grants from NSERC of
Canada and FQRNT of Qu\'ebec.

\appendix

\section{Brane-Antibrane Potential in Warped Throat}
\begin{figure}[htbp]
\begin{center}
\includegraphics[width=0.4\textwidth]{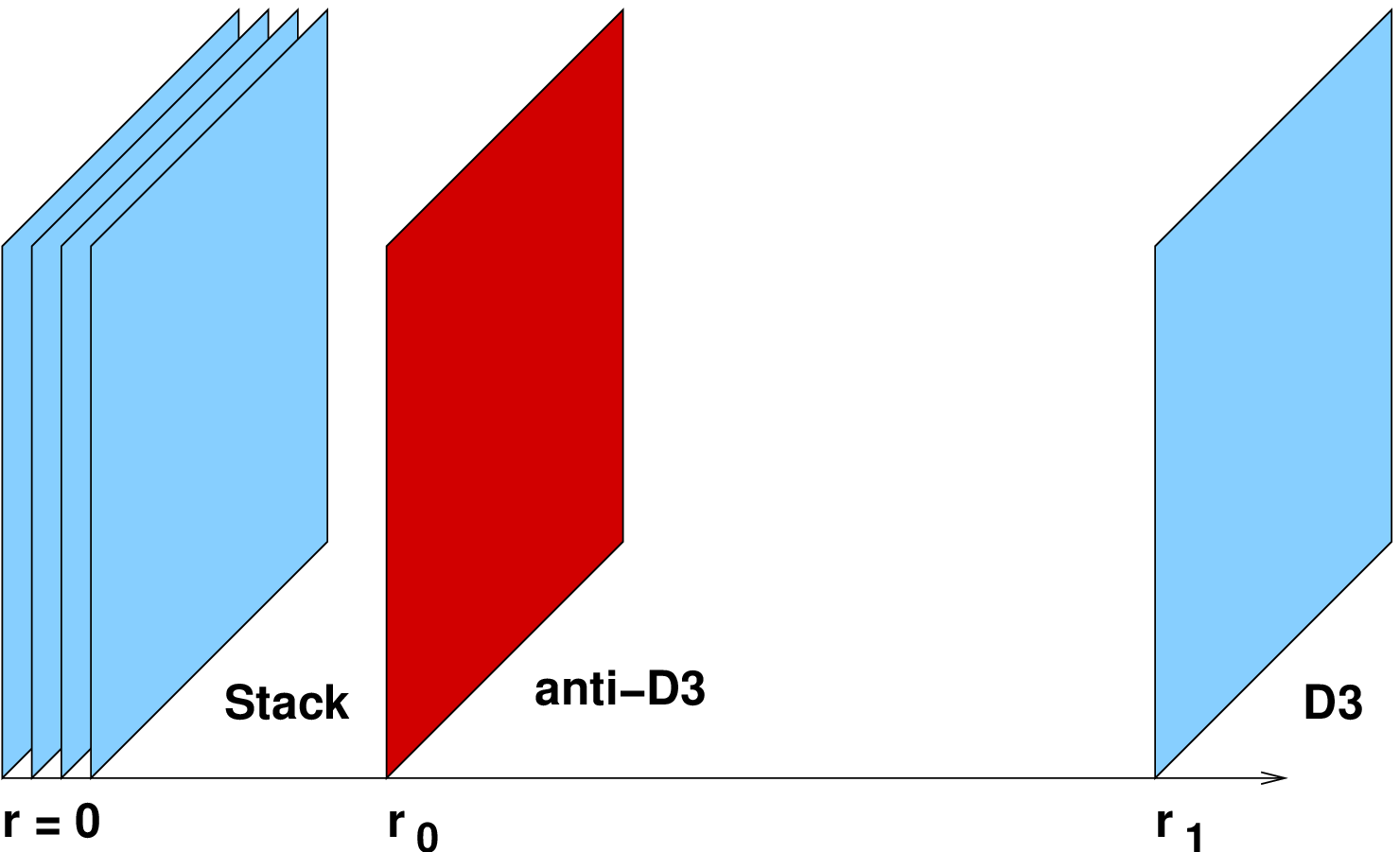}
\end{center}
\caption{\KKLMMT\ setup \label{KKLMMT_setup}}
\end{figure}

We consider the background of
$AdS_5\times X_5$ space with the $AdS_5$ metric:
\be
ds_5^2 = \frac{r^2}{R^2}\left(-dt^2+d{\bf x}^2\right) + \frac{R^2}{r^2}dr^2
\ee
where $R^4 = 4\pi a g_s N {\alpha^{\prime}}^2$ is the radius of the $AdS_5$ space
created as the near-horizon geometry of a stack of $N \gg 1$ $D3$ branes. 
There is also an $RR$ background field created by the stack of branes:
\be
A_{x^0, x^1, x^2, x^3} = 
\lim_{r\rightarrow 0}\frac{1}{1+\left(\frac{R}{r}\right)^4}\simeq
\left(\frac{r}{R}\right)^4
\ee
If one now writes the action for a probe brane is this background, one finds:
\beqa
S_{D3} &=& -T_3\int d^4\xi
\sqrt{-\det\left(\partial_{\xi^a}x^M\partial_{\xi^b}x^Ng_{NM}\right)}
\nonumber\\ &+&  T_3\int d^4\xi A_{x^0, x^1, x^2, x^3}
\eeqa
We will use the convention that $\xi^0 = t$, but the rest of the coordinates will 
describe the location of the brane, including fluctuations (local bendings), 
$x^i = x^i\left(\xi^0,\xi^i\right) = x^i\left(t,\xi^i\right)$. We make here a 
further assumption, namely that only its radial location varies. 
\begin{widetext}
\be
\partial_{\xi^a}x^M\partial_{\xi^b}x^Ng_{NM} = 
\partial_{\xi^a}x^0\partial_{\xi^b}x^0g_{00} + 
\partial_{\xi^a}x^i\partial_{\xi^b}x^ig_{ii} +
\partial_{\xi^a}r\partial_{\xi^b}rg_{rr}
\ee
The metric will look like:
\be
\left(\begin{array}{cccc}
g_{00} + g_{rr}\left(\frac{\partial r}{\partial \xi^0}\right)^2 & 
g_{rr}\frac{\partial r}{\partial \xi^0}\frac{\partial r}{\partial \xi^1} & 
g_{rr}\frac{\partial r}{\partial \xi^0}\frac{\partial r}{\partial \xi^2} & 
g_{rr}\frac{\partial r}{\partial \xi^0}\frac{\partial r}{\partial \xi^3}   \\
g_{rr}\frac{\partial r}{\partial \xi^1}\frac{\partial r}{\partial \xi^0} & 
g_{xx} + g_{rr}\left(\frac{\partial r}{\partial \xi^1}\right)^2 & 
g_{rr}\frac{\partial r}{\partial \xi^1}\frac{\partial r}{\partial \xi^2} & 
g_{rr}\frac{\partial r}{\partial \xi^1}\frac{\partial r}{\partial \xi^3}\\
g_{rr}\frac{\partial r}{\partial \xi^2}\frac{\partial r}{\partial \xi^0} & 
g_{rr}\frac{\partial r}{\partial \xi^2}\frac{\partial r}{\partial \xi^1} & 
g_{yy} + g_{rr}\left(\frac{\partial r}{\partial \xi^2}\right)^2 & 
g_{rr}\frac{\partial r}{\partial \xi^2}\frac{\partial r}{\partial \xi^3}\\
g_{rr}\frac{\partial r}{\partial \xi^3}\frac{\partial r}{\partial \xi^0} & 
g_{rr}\frac{\partial r}{\partial \xi^3}\frac{\partial r}{\partial \xi^1} & 
g_{rr}\frac{\partial r}{\partial \xi^3}\frac{\partial r}{\partial \xi^2} & 
g_{zz} + g_{rr}\left(\frac{\partial r}{\partial \xi^3}\right)^2
\end{array}\right) = 
%\nonumber \\
\left(\begin{array}{cccc}
g_{00} & 0 & 0 & 0\\
0 & g_{xx} & 0 & 0\\
0 & 0 & g_{yy} & 0\\
0 & 0 & 0 & g_{zz}
\end{array}\right) + 
\left(g_{rr}\frac{\partial r}{\partial \xi^{\mu}}
\frac{\partial r}{\partial \xi^{\nu}}\right)
\ee
We can apply the usual formula:
\be
\det\left(A+\epsilon B\right) = 
\det A\det\left(1+\epsilon A^{-1}B\right)
\simeq \det A \left(1+\epsilon\; Tr\left( A^{-1}B\right)\right)
\ee
Writing the metric at the location of the brane as:
$g_{\mu\nu} = \frac{r^2}{R^2}f_{\mu\nu}$ we obtain:
\be
\det\left(\partial_{\xi^a}x^M\partial_{\xi^b}x^Ng_{NM}\right)=
-\left(\frac{r}{R}\right)^8\left(1-\left(\frac{R}{r}\right)^4\!\!
f^{\mu\nu}
\frac{\partial r}{\partial \xi^{\mu}}\frac{\partial r}{\partial \xi^{\nu}}\right)
\ee
\end{widetext}
The factor $\frac{r^4}{R^4}$ in the bracket comes from the product 
\be
g^{\mu\nu}g_{rr} = \left(\frac{R}{r}\right)^4f^{\mu\nu}
\ee
Using these results in the $DBI+CS$ action we obtain:
\beqa
\label{DBI+CS}
S_{D3} &=& -T_3\int d^4\xi\sqrt{-f}\left(\frac{r}{R}\right)^4
\sqrt{1-\left(\frac{R}{r}\right)^4\!\!\!f^{\mu\nu}
\frac{\partial r}{\partial \xi^{\mu}}\frac{\partial r}{\partial \xi^{\nu}}}
\nonumber\\
&\pm& T_3\int d^4\xi\det\left(\frac{\partial x}{\partial \xi}\right)
\left(\frac{r}{R}\right)^4\epsilon_{t\xi^1\xi^2\xi^3}
\eeqa
Expanding the square root term we obtain for a $D3$ brane (sign $+$):
\be 
S_{D3} = T_3\int d^4\xi\sqrt{-f} \hf f^{\mu\nu}
\frac{\partial r}{\partial \xi^{\mu}}\frac{\partial r}{\partial \xi^{\nu}}
\ee
while for a $\overline{D}3$ (sign $-$):
\beqa
\label{D3_bar_action}
S_{\overline{D}3} &=& T_3\int d^4\xi\sqrt{-f} \hf f^{\mu\nu}
\frac{\partial r}{\partial \xi^{\mu}}
\frac{\partial r}{\partial \xi^{\nu}} \nonumber\\ &-& 
2T_3\int d^4\xi\sqrt{-f}\left(\frac{r}{R}\right)^4
\eeqa
The extra piece in the case of the $\overline{D}3$ comes from the non-BPS
nature of the antibrane, since the graviton exchange and the RR exchange 
forces now add instead of canceling (for 3-branes the dilaton is constant 
and contributes no force). The potential energy for the 
$\overline{D}3$ in the gravitational + RR field will make it favorable 
for the brane to move towards $r=0$, that is, the bottom of the $AdS_5$ throat.
For the setup in fig.(\ref{KKLMMT_setup}) we have to add the piece
corresponding to the graviton + RR exchange between the $D3$ 
and the $\overline{D}3$.
However in this setup we have to include the effect of the $D3$ in addition 
to that of the stack. The derivation of the background perturbed by the 
presence of the $D3$ placed away from the stack can be found in
Appendix B of \KKLMMT.
It starts by writing the metric in the form:
\beqa
ds_{10}^2 &=& h^{-\hf}\left(-dt^2+d{\bf x}^2\right)+
h^{\hf}\left(dr^2+\frac{r^2}{R^2}\overline{g}_{ab}dy^ady^b\right)
\nonumber\\ h&=&\frac{R^4}{r^4}
\eeqa
placing the extra $D3$ brane at $r_1$ and solving for the perturbed harmonic function
\be
h\left(r\right)=\frac{R^4}{r^4} + \delta h\left(r\right)
\ee
which satisfies the Laplace equation in the space transverse to the brane, which is 
spanned by $r$ and $y^a$:
\be
\nabla^2_{\left(6\right)}\delta h\left(r\right) = C\delta^{\left(6\right)}
\left(\vec{r} - \vec{r}_{1}\right)
\ee
which amounts to solving:
\be
\frac{R^5}{r^5}\partial_r\left[\frac{r^5}{R^5}\partial_r \delta h\left(r\right)\right] = 
C\delta^{\left(6\right)}\left(\vec{r} - \vec{r}_{1}\right)
\ee
which give $\delta h\left(r\right)\sim |\vec r-\vec r_1|^{-4}$.
A similar equation is satisfied by the unperturbed $h$.
The constants $C$ and $C^{\prime}$ depend on the total ``charge'', more precisely 
the total tension of the brane or stack of branes, so they are related by:
\be
C = \frac{C^{\prime}}{N}
\ee
and since the solution for the unperturbed $ h\left(r\right)$ is already known, 
we can immediately find the solution for $ \delta h\left(r\right)$ 
\be
\delta h\left(r\right) = \frac{C}{|\vec r - \vec r_1|^4} 
= \frac{R^4}{N}\cdot\frac{1}{|\vec r - \vec r_1|^4}
\ee
We can use the perturbed metric to calculate the action for
the $\overline{D}3$ using the new function $h\left(r\right) + \delta h\left(r\right)$.

\begin{figure}[h]
\begin{center}
\includegraphics[width=0.4\textwidth]{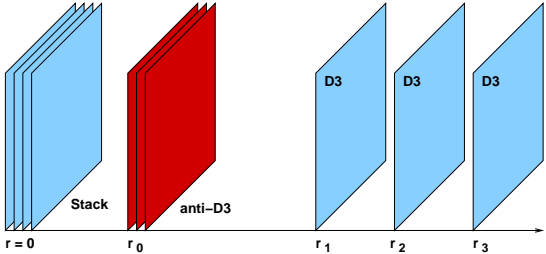}
\end{center}
\caption{Modified setup \label{modified_setup}}
\end{figure}

Now modify the setup and place at $r_0$ a stack of $M$ 
$\overline{D}3$'s with
$M \ll N$ and an equal number of $D3$'s placed at 
$r_1, r_2 \dots r_M$, as shown in figure \ref{modified_setup}.
We can simply generalize the analysis above to come to the result:
\be
h\left(r\right) = \frac{R^4}{r^4} + 
\frac{R^4}{N}\cdot\sum_{j=1}^{M}\frac{1}{|\vec r - \vec r_j|^4}
\ee
There is no interaction between the $D3$'s or between the $\overline{D}3$'s in 
the second stack since they are mutually BPS.
We can now go back to eq.\ (\ref{DBI+CS}) and replace the expression 
$R^4/r^4$ by $h\left(r\right)$.
The relevant term is the second term from eq.\ (\ref{D3_bar_action}), 
which in the limit $r_i\gg r$ becomes:
\begin{widetext}
\be
2T_3\int d^4\xi\sqrt{-f}\frac{1}{h\left(r\right)} = 
2T_3\int d^4\xi\sqrt{-f}\frac{r_0^4}{R^4}\cdot
\frac{1}{1+\frac{r_0^4}{N}\sum_{j=1}^{M}\frac{1}{r_j^4}}
%\nonumber \\
\cong\int d^4\xi\sqrt{-f}\left\{2T_3\cdot\frac{r_0^4}{R^4}\cdot
\left(1-\frac{1}{N}\sum_{j=1}^{M}\frac{r_0^4}{r_j^4}\right)\right\}
\ee
\end{widetext}
where $r_0$ is the location of the stack of antibranes. The above result 
is for one single $\overline{D}3$ in the stack. The final form of the 
effective action should be multiplied by $M$, the number of antibranes in
the stack.

\section{K\"ahler potential corrections}

The K\"ahler potential is assumed to have the following form:
\be
K=-3\log\left(r\right)=
-3\log\left(T+\overline{T}-f\left(\phi_i,\overline{\phi}_i\right)\right)
\ee
and we will also assume that $f$ decomposes into a sum of functions,
each one depending on one field $\phi_i$:
\be
\label{KPot}
K=-3\log\left(T+\overline{T}-\sum_{i}f_i\left(\phi_i,\overline{\phi}_i\right)\right)
\ee
The simplest form of the functions $f_i$ is
\be
f_i\left(\phi_i,\overline{\phi}_i\right) = \phi_i\overline{\phi}_i
\ee 
but we can consider more complicated expressions, for example:
\be
f_i\left(\phi_i,\overline{\phi}_i\right) = \phi_i\overline{\phi}_i -
\left(\phi_i\overline{\phi}_i\right)^2
\ee

Allowing for the more complicated expression for $f_i\left(\phi_i,\overline{\phi}_i\right)$
changes the structure of the singularities of the K\"ahler potential and of the
F-term. 
%In the simplest case they were the same. For general $f_i$'s
%the F-term becomes:
%\be
%V_F\sim\frac{\text{big mess}}{r^3\cdot
%\frac{\partial^2 f_1}{\partial\phi_1\partial\overline{\phi}_1}
%\frac{\partial^2 f_2}{\partial\phi_2\partial\overline{\phi}_2}
%\cdots
%\frac{\partial^2 f_n}{\partial\phi_n\partial\overline{\phi}_n}}
%\ee
%In the original scenario the derivatives were trivial, 
%$\frac{\partial^2 f_i}{\partial\phi_i\partial\overline{\phi}_i}=1$, and did not 
%introduce additional singularities. In the case 
%\be
%\label{non_min}
%f_i\left(\phi_i,\overline{\phi}_i\right) = \phi_i\overline{\phi}_i -
%\left(\phi_i\overline{\phi}_i\right)^2
%\ee
%we get new singularities at $\left|\phi_i\right|=\hf$.

%%%%%%%%%%%%%%%%%%%%%%%%%%%%%%%%%%%%%%%%%%%%%%%%%%%%%%%%%%%%%%%%%%%%%%%%%%%%%%%%%%%%
In the general case the K\"ahler metric derived  from the K\"ahler
potential of eq.\ (\ref{KPot}) 
has the form:
\begin{widetext}
\be
\label{Kahler_Metric}
{\mathcal K}_{a\overline{b}} = 
\frac{3}{r^2}\left(\begin{array}{ccccc}
1 & -\frac{\partial f_1}{\partial\phi_1} & -\frac{\partial f_2}{\partial\phi_2} & 
\cdots & -\frac{\partial f_n}{\partial\phi_n} \\
-\frac{\partial f_1}{\partial\phi_1^*} & 
\frac{\partial f_1}{\partial\phi_1}\frac{\partial f_1}{\partial\phi_1^*}+ 
r\frac{\partial^2 f_1}{\partial\phi_1\partial\phi_1^*} &
\frac{\partial f_1}{\partial\phi_1^*}\frac{\partial f_2}{\partial\phi_2} & 
\cdots & \frac{\partial f_1}{\partial\phi_1^*}\frac{\partial f_n}{\partial\phi_n} \\
\vdots & \vdots & \vdots & \ddots & \vdots \\
-\frac{\partial f_{n-1}}{\partial\phi_{n-1}^*} 
&\frac{\partial f_{n-1}}{\partial\phi_{n-1}^*}\frac{\partial f_1}{\partial\phi_1}  
&\frac{\partial f_{n-1}}{\partial\phi_{n-1}^*}\frac{\partial f_2}{\partial\phi_2} 
&\cdots 
&\frac{\partial f_{n-1}}{\partial\phi_{n-1}^*}\frac{\partial f_n}{\partial\phi_n} \\
-\frac{\partial f_{n}}{\partial\phi_{n}^*} 
&\frac{\partial f_n}{\partial\phi_n^*}\frac{\partial f_1}{\partial\phi_1}
&\frac{\partial f_n}{\partial\phi_n^*}\frac{\partial f_2}{\partial\phi_2}
&\cdots 
&\frac{\partial f_n}{\partial\phi_n^*}\frac{\partial f_n}{\partial\phi_n}+
r\frac{\partial^2 f_n}{\partial\phi_n\partial\phi_n^*}
\end{array}\right)
\ee
The inverse can also be calculated:
\be
\label{Kmetric}
{\mathcal K}^{\overline{a}b} = 
\frac{r}{3}\left(\begin{array}{ccccc}
r+\frac{\partial f_1}{\partial\phi_1}\frac{\partial f_1}{\partial\phi_1^*}/
\frac{\partial^2 f_1}{\partial\phi_1\partial\phi_1^*} + \dots +
\frac{\partial f_n}{\partial\phi_n}\frac{\partial f_n}{\partial\phi_n^*}/
\frac{\partial^2 f_n}{\partial\phi_n\partial\phi_n^*} & 
\frac{\partial f_1}{\partial\phi_1^*}/\frac{\partial^2 f_1}{\partial\phi_1\partial\phi_1^*}& 
\frac{\partial f_2}{\partial\phi_2^*}/\frac{\partial^2 f_2}{\partial\phi_2\partial\phi_2^*}& 
\hdots & 
\frac{\partial f_n}{\partial\phi_n^*}/\frac{\partial^2 f_n}{\partial\phi_n\partial\phi_n^*}\\
\frac{\partial f_1}{\partial\phi_1}/\frac{\partial^2 f_1}{\partial\phi_1\partial\phi_1^*}& 
1/\frac{\partial^2 f_1}{\partial\phi_1\partial\phi_1^*} & 0& \hdots & 0\\
\frac{\partial f_2}{\partial\phi_2}/\frac{\partial^2 f_2}{\partial\phi_2\partial\phi_2^*} & 
0 &
1/\frac{\partial^2 f_2}{\partial\phi_2\partial\phi_2^*} & \hdots & 0\\
\vdots & \vdots &\vdots &\ddots &\vdots \\ 
\frac{\partial f_n}{\partial\phi_n}/\frac{\partial^2 f_n}{\partial\phi_n\partial\phi_n^*} & 
0& 0& \hdots & 1/\frac{\partial^2 f_n}{\partial\phi_n\partial\phi_n^*}
\end{array}\right)
\ee
\end{widetext}

The F-term is calculated using the usual formula:
\be
V_F=e^{\mathcal K}\left(K^{\overline{a}b}\overline{D_{a}W}D_{b}W-3\left|W\right|^2\right)
\ee
where 
\be
{\mathcal K} = -3\log\left(r\right)
\ee
The covariant derivatives are expressed as:
\baray
D_{T}W &=& \partial_TW+W\partial_T{\mathcal K} = \partial_TW-\frac{3}{r}W\\
D_{\phi_i}W &=& \partial_{\phi_i}W+W\partial_{\phi_i}{\mathcal K} = 
\frac{\partial f_i}{\partial\phi_i}\frac{3}{r}W\\
\overline{D_{T}W} &=& \partial_{\overline{T}}\overline{W}+
\overline{W}\partial_{\overline{T}}{\mathcal K} = 
\partial_{\overline{T}}\overline{W}-\frac{3}{r}\overline{W}\\
\overline{D_{\phi_i}W} &=& \partial_{\overline{\phi_i}}\overline{W}+
\overline{W}\partial_{\overline{\phi_i}}{\mathcal K} = 
\frac{\partial f_i}{\partial\phi_i^*}\frac{3}{r}\overline{W}
\earay
\begin{widetext}
Using the above expression for the inverse K\"ahler metric one obtains the following 
terms:
\baray
K^{\overline{T}T}\overline{D_{T}W}D_{T}W &=& \frac{r}{3}
\left(\partial_{\overline{T}}\overline{W}-\frac{3}{r}\overline{W}\right)
\left(\partial_TW-\frac{3}{r}W\right)
\left(r+\frac{\frac{\partial f_1}{\partial\phi_1}\frac{\partial f_1}{\partial\phi_1^*}}
{\frac{\partial^2 f_1}{\partial\phi_1\partial\phi_1^*}} + \dots +
\frac{\frac{\partial f_n}{\partial\phi_n}\frac{\partial f_n}{\partial\phi_n^*}}
{\frac{\partial^2 f_n}{\partial\phi_n\partial\phi_n^*}}\right) \\
K^{\overline{T}\phi_i}\overline{D_{T}W}D_{\phi_i}W &=& \frac{r}{3}
\left(\partial_{\overline{T}}\overline{W}-\frac{3}{r}\overline{W}\right)W
\left(\frac{3}{r}\frac{\partial f_1}{\partial\phi_1} \frac{\frac{\partial f_1}{\partial\phi_1^*}}
{\frac{\partial^2 f_1}{\partial\phi_1\partial\phi_1^*}}
+ \dots +
\frac{3}{r}\frac{\partial f_n}{\partial\phi_n}\frac{\frac{\partial f_n}{\partial\phi_n^*}}
{\frac{\partial^2 f_n}{\partial\phi_n\partial\phi_n^*}}\right)\\
K^{\overline{\phi_i}T}\overline{D_{\phi_i}W}D_{T}W &=& \frac{r}{3}
\left(\partial_TW-\frac{3}{r}W\right)\overline{W}
\left(\frac{3}{r}\frac{\partial f_1}{\partial\phi_1^*} \frac{\frac{\partial f_1}{\partial\phi_1}}
{\frac{\partial^2 f_1}{\partial\phi_1\partial\phi_1^*}}
+ \dots +
\frac{3}{r}\frac{\partial f_n}{\partial\phi_n^*}\frac{\frac{\partial f_n}{\partial\phi_n}}
{\frac{\partial^2 f_n}{\partial\phi_n\partial\phi_n^*}}\right)\\
K^{\overline{\phi_i}\phi_j}\overline{D_{\phi_i}W}D_{\phi_j}W &=& \frac{r}{3}W\overline{W}
\left(\frac{9}{r^2}\frac{\partial f_1}{\partial\phi_1}\frac{\partial f_1}{\partial\phi_1^*}
\frac{1}
{\frac{\partial^2 f_1}{\partial\phi_1\partial\phi_1^*}} + \dots +
\frac{9}{r^2}\frac{\partial f_n}{\partial\phi_n}\frac{\partial f_n}{\partial\phi_n^*}\frac{1}
{\frac{\partial^2 f_n}{\partial\phi_n\partial\phi_n^*}}\right)
\earay

Adding all the above terms and $-3\left|W\right|^2$ we obtain the F-term:
\be
V_F=\frac{1}{r^3}\left[\frac{r^2}{3}\partial_{\overline{T}}\overline{W}\partial_TW
\left(1+\frac{\frac{\partial f_1}{\partial\phi_1}\frac{\partial f_1}{\partial\phi_1^*}}
{r\frac{\partial^2 f_1}{\partial\phi_1\partial\phi_1^*}} + \dots +
\frac{\frac{\partial f_n}{\partial\phi_n}\frac{\partial f_n}{\partial\phi_n^*}}
{r\frac{\partial^2 f_n}{\partial\phi_n\partial\phi_n^*}}\right)-
r\left(W\partial_{\overline{T}}\overline{W} + 
\overline{W}\partial_TW\right)\right]
\ee
\end{widetext}
We assumed that the superpotential itself is independent of the fields 
$\phi_1 \dots \phi_n$.
%\end{widetext}

\end{document}